\PassOptionsToPackage{table}{xcolor}
\documentclass[sigconf]{acmart}

\usepackage{listings} 
\usepackage{xcolor}
\definecolor{codered}{rgb}{0.6,0,0}
\definecolor{codegreen}{rgb}{0,0.6,0}
\definecolor{codegray}{rgb}{0.5,0.5,0.5}
\definecolor{codepurple}{rgb}{0.58,0,0.82}
\definecolor{backcolour}{rgb}{0.95,0.95,0.92}
\lstdefinestyle{mystyle}{
    backgroundcolor=\color{backcolour},   
    commentstyle=\color{codegreen},
    keywordstyle=\color{magenta},
    numberstyle=\tiny\color{codegray},
    stringstyle=\color{codepurple},
    basicstyle=\ttfamily\footnotesize,
    breakatwhitespace=false,         
    breaklines=true,                 
    captionpos=b,                    
    keepspaces=true,                 
    numbersep=5pt,                  
    showspaces=false,                
    showstringspaces=false,
    showtabs=false,                  
    tabsize=2
}
\usepackage{enumitem}
\usepackage{tikz}
\definecolor{customMagenta}{HTML}{E52CCF}

\usepackage{tabularx}   
\usepackage{booktabs}  
\usepackage{array} 
\AtBeginDocument{%
  \providecommand\BibTeX{{%
    \normalfont B\kern-0.45em{\scshape i\kern-0.25em b}\kern-0.8em\TeX}}}

\copyrightyear{2025}
\acmYear{2025}
\setcopyright{cc}
\setcctype{by}
\acmConference[Koli Calling 2025]{In 25th Koli Calling International Conference on Computing Education Research (Koli Calling '25),}{November 11-November 16, 2025}{Koli, Finland}
\acmBooktitle{In 25th Koli Calling International Conference on Computing Education Research (Koli Calling '25), November  11-16, Koli}\acmDOI{}
\acmISBN{}

\copyrightyear{2025}
\acmYear{2025}
\setcopyright{cc}
\setcctype{by}
\acmConference[Koli Calling '25]{25th Koli Calling International Conference on Computing Education Research}{November 11--16, 2025}{Koli, Finland}
\acmBooktitle{25th Koli Calling International Conference on Computing Education Research (Koli Calling '25), November 11--16, 2025, Koli, Finland}
\acmPrice{}
\acmDOI{10.1145/3769994.3770036}
\acmISBN{979-8-4007-1599-0/25/11}

\begin{document}


\keywords{literature review, large language models, harms, computing education}



\title[Beyond the Benefits]{Beyond the Benefits: A Systematic Review of Risks, Harms, and Unintended Consequences of Generative AI in Computing Education}

\title[Beyond the Benefits]{Beyond the Benefits: A Systematic Review of the Harms and Unintended Consequences of Generative AI in Computing Education}

    \title[Beyond the Benefits]{Beyond the Benefits: A Systematic Review of the Harms and Consequences of Generative AI in Computing Education}




\author{Seth Bernstein}
\affiliation{%
  \institution{Temple University}
  \city{Philadelphia, PA}
  \country{USA}
}
\email{seth.bernstein@temple.edu}
\orcid{0000-0002-7552-5448}

\author{Ashfin Rahman}
\affiliation{%
  \institution{Temple University}
  \city{Philadelphia, PA}
  \country{USA}
}
\email{ashfin.rahman@temple.edu}
\orcid{0009-0003-8400-8359}

\author{Nadia Sharifi}
\affiliation{%
  \institution{Temple University}
  \city{Philadelphia, PA}
  \country{USA}
}
\email{nadia.sharifi@temple.edu}
\orcid{0009-0009-9055-0332}

\author{Ariunjargal Terbish}
\affiliation{%
  \institution{Temple University}
  \city{Philadelphia, PA}
  \country{USA}
}
\email{ariunjargal.terbish@temple.edu}
\orcid{0009-0000-1829-6231}

\author{Stephen MacNeil}
\affiliation{%
  \institution{Temple University}
  \city{Philadelphia, PA}
  \country{USA}
}
\email{stephen.macneil@temple.edu}
\orcid{0000-0003-2781-6619}


\renewcommand{\shortauthors}{Bernstein et al.}

\begin{abstract}

Generative artificial intelligence (GenAI) has already had a big impact on computing education with prior research identifying many benefits. However, recent studies have also identified potential risks and harms. To continue maximizing AI benefits while addressing the harms and unintended consequences, we conducted a systematic literature review of research focusing on the risks, harms, and unintended consequences of GenAI in computing education. Our search of ACM DL, IEEE Xplore, and Scopus (2022-2025) resulted in 1,677 papers, which were then filtered to 224 based on our inclusion and exclusion criteria. Guided by best practices for systematic reviews, four reviewers independently extracted publication year, learner population, research method, contribution type, GenAI technology, and educational task information from each paper. We then coded each paper for concrete harm categories such as academic integrity, cognitive effects, and trust issues. Our analysis shows patterns in how and where harms appear, highlights methodological gaps and opportunities for more rigorous evidence, and identifies under-explored harms and student populations. By synthesizing these insights, we intend to equip educators, computing students, researchers, and developers with a clear picture of the harms associated with GenAI in computing education.

\end{abstract}

\keywords{}

\ccsdesc[500]{Social and professional topics~Computing education}

\keywords{large language models, generative AI, harms, computing education}

\maketitle


\section{Introduction}


Over the last few years, generative AI (GenAI) has had a transformative impact on computing education~\cite{prather2023robots, prather2024beyond}. For example, GenAI can explain code to students better than their peers~\cite{macneil2023experiences,leinonen2023comparing}, can provide hints to help students get unstuck~\cite{kiesler2023exploring, roest2024next}, and can make learning materials more personally or culturally relevant~\cite{bernstein2024like}. Classes have adopted AI-first approaches that introduce the most engaging parts of the computing curriculum earlier and delay more tedious or frustrating aspects of early programming, such as syntax and boilerplate code~\cite{porter2024learn, vadaparty2024cs1, reeves2024prompts}. Tools are also being developed to more carefully scaffold the use of generative AI in classrooms~\cite{denny2024desirable, kazemitabaar2024codeaid, liffiton2024codehelp,  denny2024prompt, macneil2025fostering}. For example, Denny et al. introduced Prompt Problems as a technique designed to help students practice prompt engineering by automatically generating and evaluating the resulting code~\cite{denny2024prompt}. MacNeil et al. introduced Autocompletion Quizzes to give students practice differentiating between good and bad code suggestions~\cite{macneil2025fostering}. These two examples show how pedagogical tools can not only facilitate learning, but actually teach new skills and competencies that students need to use AI effectively.  

Alongside these benefits and innovative tools and pedagogies, researchers have also raised concerns about negative impacts on learning and unintended consequences. Early concerns included students over-relying on GenAI~\cite{zastudil2023generative, becker2023programming, lau2023from}, academic dishonesty~\cite{denny2024computing, hou2024more, gutierrez2024seeing, savelka2023thrilled, zastudil2023generative}, and uneven benefits, with the best students receiving more value from AI~\cite{zastudil2023generative, hou2024effects, prather2024widening}. Some have even compared this to the `digital divide' of the 1990s~\cite{zastudil2023generative}. Subsequent empirical studies have provided evidence that support some these concerns. For example, GenAI models can solve problems at the CS1 and CS2 levels~\cite{finnie-ansley2023my, savelka2023thrilled}, Parsons problems~\cite{reeves2023evaluating, hou2024more}, and graph and tree data structure tasks using only image-based inputs~\cite{gutierrez2024seeing}. Recent work has also provided empirical evidence for a widening gap between students who can use AI effectively and those who struggle~\cite{prather2024widening, margulieux2024self}. A recent study by Hou et al. also showed negative impacts on peer interactions and learning communities~\cite{hou2025all}.


These negative impacts are concerning, but they are not yet well-understood. To address this gap, we conducted a systematic literature review of studies on GenAI in computing education that discuss harms, challenges, and unintended consequences. Our review focuses on research published between 2022 and 2025. In this review, we define `harms' broadly to include cognitive impacts (e.g., over-reliance), ethical risks (e.g., plagiarism), social and equity concerns, and unintended pedagogical effects. Our goal was to investigate the following research questions:

\begin{itemize}
    \item[\textbf{RQ1:}] What types of harms or unintended consequences of generative AI have been reported? \textbf{(Types of Harms)}  
    \item [\textbf{RQ2:}] What empirical evidence supports each identified harm? \textbf{(Evidence)} 
    \item[\textbf{RQ3:}] In what educational contexts have these harms been examined or discussed? \textbf{(Contexts)} 
\end{itemize}

Our search resulted in 1,849 references from the three popular computer science databases: ACM Digital Library, IEEE Xplore, and Scopus. After removing duplicate and retracted papers, 1,677 references remained. After filtering these papers based on our inclusion and exclusion criteria, we had 224 total references which were analyzed in detail. We evaluated the types of harms and their prevalence across various contexts including GenAI tool types and populations. We also coded for research contributions, methods, and levels of evidence.

Our findings show a few clear takeaways: (1) researchers are increasingly attending to the potential harms of generative AI, (2) many of these harms remain insufficiently supported by empirical evidence, and (3) certain categories of harm such as those related to social interactions and collaboration are still under-represented in the literature. This paper makes two primary contributions: (1) a taxonomy of generative AI harms specific to computing education, and (2) empirical insights into how these harms are currently framed and addressed in the field.

\begin{figure*}
    \centering
    \includegraphics[width=0.85\linewidth]{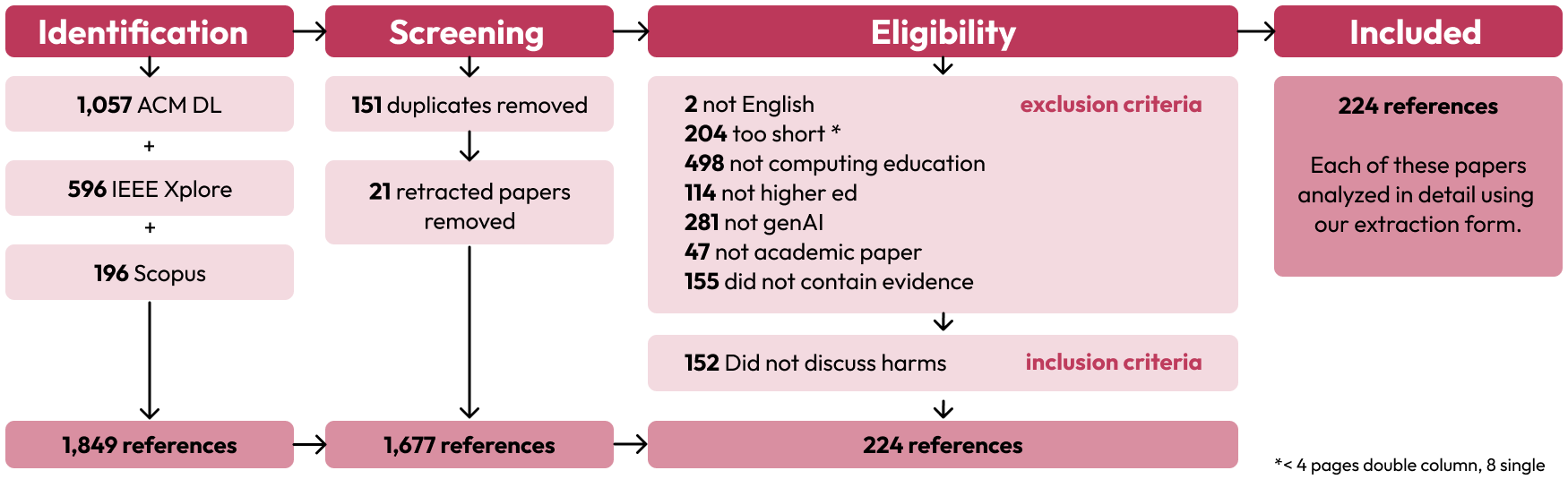}
    \caption{PRISMA diagram detailing our identification, screening, and assessment of paper eligibility for inclusion.}
    \label{fig:prismaslrharms}
\end{figure*}

\section{Methodology}

The goal of this research is to understand the negative impacts and unintended consequences of GenAI in computing education. We conducted a systematic literature review (SLR) following the Preferred Reporting Items for Systematic Reviews and Meta-analysis (PRISMA) reporting guidelines~\cite{page2021prisma}. Below we discuss the details of this process, which are also summarized in Figure~\ref{fig:prismaslrharms}.


\subsection{Scope of the Review}

To ensure a focused and manageable review, we first define the scope of our search, including inclusion/exclusion criteria and temporal boundaries.

\subsubsection{Inclusion and Exclusion Criteria}

Over the last few years there has been a substantial increase in the number of GenAI papers~\cite{prather2023robots,prather2024beyond}. We adopted multiple exclusion criteria which are summarized in Table~\ref{tab:excl-incl} to help us focus on the most relevant papers. We excluded papers that were not written in English, that did not focus on generative AI, and papers that were not specifically about computing education. To further reduce the scope, we focused on higher education and omitted k-12 research. We also excluded non-peer-reviewed materials, such as compiled conference proceedings and other informal publications.

\subsubsection{Temporal scope and year reporting}
Our searches covered 2022-2025 and concluded in July 2025. For any year-by-year plots, we intentionally omit 2025. Including 2025 would bias counts towards venues with early publication schedules (e.g., SIGCSE) and underrepresent later 2025 conferences whose proceedings publish after our cutoff (e.g., ICER, ITiCSE).



\begin{table}[h]
\small
\centering
\caption{Exclusion and inclusion criteria.}
\label{tab:excl-incl}
\begin{tabular}{@{}l@{}}
\toprule
\textbf{Exclusion Criteria} \\ \midrule
\textbf{EC$_1$:} Not available in English \\
\textbf{EC$_2$:} No clear generative AI focus \\
\textbf{EC$_3$:} Not related to computing education \\
\textbf{EC$_4$:} Focus is not about higher education (i.e: k-12) \\
\textbf{EC$_5$:} Not an academic paper (e.g.: conference proceedings) \\
\textbf{EC$_6$:} Does not contain evidence \\
\toprule
\textbf{Inclusion Criteria} \\ \midrule
\begin{tabular}[c]{@{}l@{}}\textbf{IC$_1$:} Discusses or demonstrates generative AI harms \end{tabular} \\

\end{tabular}
\end{table}

\subsection{Search and Selection Process}

As mentioned, our process is guided by the PRISMA reporting guidelines, following three phases~\cite{page2021prisma}. In the \textbf{Identification Phase}, we identified potentially qualifying citations by searching the ACM Digital Library, IEEE Xplore, and Scopus. During the \textbf{Screening Phase}, we filtered the resulting papers to ensure papers meeting the exclusion criteria were rejected. With the \textbf{Eligibility Phase}, we further filtered the set of references resulting from the screening phase to verify that the inclusion criteria was met.

\subsubsection{Identification Phase}

To ensure an overarching and comprehensive sample, we used three databases in our literature review: ACM Digital Library, IEEE Xplore, and Scopus. These databases were selected to align with those used in recent working group reviews on generative AI in computing education~\cite{prather2024beyond, prather2023robots}. ACM DL and IEEE Xplore include publications from prominent computing education conferences, such as SIGCSE Technical Symposium, ITiCSE, and FIE. These databases also include publications from ACM TOCE and other relevant journals. Scopus includes some overlap with both databases while also including some relevant journals that the other databases do not include. 


\begin{itemize}
    \item \texttt{\textbf{Domain:}} ``computer science education'' OR ``computing education'' OR ``CS education'' OR ``CSEd'' OR ``CER'' OR ``programming education'' OR ``introductory programming'' OR ``CS1'' OR ``CS2'' OR ``computing student*'' OR ``CS student*'' OR ``computer science student*''
    \item \texttt{\textbf{Topic:}} ``LLM*'' OR ``large language model*'' OR ``generative AI'' OR ``ChatGPT'' OR ``GPT'' OR ``Bard'' OR ``Gemini'' OR ``Claude'' OR ``GenAI'' OR ``Copilot'' OR ``artificial intelligence'' OR ``AI''
    \item \texttt{\textbf{Publication Date:}} Jan 2022 to Jul 2025 
\end{itemize}


We combined these terms to find papers that focus on computing education and also involve large language models or generative AI.

\begin{quote}
\texttt{\textbf{Search String}} = \texttt{\textbf{Domain}} AND \texttt{\textbf{Topic}}      
\end{quote}



\subsubsection{Screening Phase}

As a first step in filtering the papers, we removed 151 duplicate papers and 21 papers that had been marked as `retracted' by the conference or journal. After removing these 172 papers, we were left with 1,677 references.  

\subsubsection{Eligibility Phase}

In the second filtering pass, papers were excluded if they did not meet the inclusion criteria or violated any exclusion criteria (Table~\ref{tab:excl-incl}). Four researchers conducted this in two stages. In the first stage, the team applied the exclusion criteria, achieving an inter-rater reliability (IRR) of 0.84 using Fleiss' Kappa. In the second stage, papers were retained only if they met the inclusion criteria, with an IRR of 0.69. This process resulted in 224 papers retained and 1,453 papers removed.

\subsection{Data Analysis}


We conducted a deductive coding of each included study across several dimensions, drawing on established taxonomies where possible. The dimensions include contribution type, research methods, type of generative AI tool examined, and participant population. 



\subsubsection{Research Methods}

We coded each study based on the reported research methods using a modified version of the taxonomy by Mack et al.~\cite{mack2021we}. Their taxonomy includes \textbf{controlled experiment}, \textbf{usability study}, \textbf{field study}, \textbf{interview}, \textbf{questionnaire}, \textbf{case study}, \textbf{focus group}, \textbf{workshop or design study}, \textbf{randomized control trials}. Based on our initial reading of the papers, we realized ethnographic methods and benchmarking were also common and we added those codes as well.

Studies could be coded as having multiple research methods to account for mixed-methods research. For example, there were multiple cases where interviews were used alongside surveys. 

\subsubsection{GenAI Tool Type}


GenAI tools continue to evolve from the early language models to conversational agents, multimodal models, and specialized pedagogical tools that have been developed to explicitly support education. To capture this diversity, we coded for the following types of GenAI tools: 

\begin{itemize}
    \item \textbf{Chat Models} --- Conversational agents such as ChatGPT, Claude, or Gemini.  
    \item \textbf{Code Assistants} --- Code-generation tools such as GitHub Copilot or Amazon CodeWhisperer.  
    \item \textbf{Search-Integrated Models} --- AI tools embedded within search engines, such as Perplexity or Gemini Search.  
    \item \textbf{Image Models} --- Generative image models such as DALL-E or Midjourney.  
    \item \textbf{Custom or Specialized Tools} --- Tools or prototypes that integrate generative AI for explicit educational goals.  
\end{itemize}

As these tools continue to evolve and improve, it is helpful to know which harms have been associated with each type of tool. For example, some students face metacognitive challenges when using Code Assistants~\cite{prather2024widening} and it is useful to understand whether and how these harms extend to other tools.  

\subsubsection{Task}

Harms depend on the context of use and some might be more heavily associated with some tasks than others. We adopted the `Context of Usage' taxonomy from Hou et al.~\cite{hou2024effects}, which describes four primary use cases for GenAI in computing education. These include \textbf{understanding course concepts}, \textbf{debugging code}, \textbf{identifying corner cases}, and \textbf{writing code}. This taxonomy was based on students perspectives, so we also included tasks from the instructors perspective including \textbf{assignment creation}, \textbf{image generation}, \textbf{data generation}, and \textbf{feedback}. These tasks were identified thematically as we read the papers during the filtering process (i.e.: screening and eligibility).  

\subsubsection{Population}

A recent working group formed to explore how GenAI is used in upper-level courses~\cite{bouvier2025GenAI}, because, as they highlight in their literature review, most research about GenAI in computing education has focused only on the introductory courses. Therefore, we coded for the following student populations:

\begin{itemize}
    \item \textbf{Introductory undergraduates} --- students enrolled in CS1 or other entry-level courses
    \item \textbf{Upper-level undergraduates} --- students enrolled in undergraduate courses beyond the introductory sequence
    \item \textbf{Graduate students}
    \item \textbf{Instructional staff} --- including instructors, teaching assistants, or course designers
    \item \textbf{Unspecified undergraduates} --- studies that involved undergraduate students but did not specify course level

\end{itemize}

\subsubsection{Levels of Evidence}

During our initial reading, we observed that many papers described harms hypothetically and few papers provided explicit evidence for the harms that they described. For example, instructors might \textbf{hypothesize} that their students would develop misconceptions when viewing incorrect code generated by an AI coding assistant. While this is a plausible concern, it is presented speculatively, without empirical evidence. In cases where evidence was provided, it ranged from informal \textbf{observations} to more rigorous forms of \textbf{measured} evidence. We adopted a broad definition of measured evidence, including not only controlled experiments but also \textbf{surveys, interviews, and benchmarking studies.} It is important to note that we explicitly coded each paper for the highest level of evidence it provided for any harm instance in the paper, and only one label was applied for each paper. 


\subsubsection{Types of Harms}

To identify the types of harms, we first familiarized ourselves with the data by open-coding~\cite{strauss2004open} a randomly selected subset of the papers. As we continued coding the data throughout the filtering process, we expanded the codebook in the rare cases where a new harm was identified. When new codes emerged, we retroactively reviewed previously coded papers to determine whether these newly identified harms were present. We began coding with a very fine-grained coding scheme to capture very specific harms. In later stages, we combined codes and established themes to fit the data. This inductive process enabled us to identify unexpected harms and we exercised reflexivity~\cite{braun2021one} throughout the process to reflect on how our positionality (see Section~\ref{sec:positionality}) might impact our analysis. These harms and sub-harms are described along with the coded papers in Table~\ref{tab:slr-harms-cog} and Table~\ref{tab:classroom-harms}.


\begin{table*}[t]
  \centering
  \caption{Cognitive and Metacognitive Harms identified in the SLR}
  \label{tab:slr-harms-cog}
  \begin{tabularx}{\textwidth}{@{}
      >{\arraybackslash}p{3cm}   
      X                                      
      >{\arraybackslash}p{9cm}  
      @{}}
    \toprule
    \textbf{Harm} & \textbf{Definition} & \textbf{Relevant Papers} \\
    \midrule
        \rowcolor{gray!10}
 \textbf{Cognitive}&
       &
       \\
    Acquiring Knowledge &
      Incorrect information (i.e., Hallucinations) or shallow learning (e.g.: copy and pasting code) &
      \footnotesize
\cite{aerts2024feasibility,ahmed2024potentiality, ai2024custom,  al2024can, al2024ai, alghamdi2024exploring, alomar2025nurturing, alves2024give,  amoozadeh2024trust, 10578649,  anderson2024using, apiola2024first, 10578934, arora2025analyzing, axelsson2024assistance, azaiz2024feedback,  azoulay2025integrity, balse2023evaluating, barambones2024ChatGPT, barlowe2024wip, bassner2024iris, basu2024large, bird2024faceless, borghoff2025generative, brockenbrough2024using, 10871836, 10871836, budhiraja2024s, buffardi2024designing, cai2024advancing, cai2024compat, cai2024compat, 10893528, 10893528, castillo2024ethical, 10706931,  10.1145/3605507.3610629, servin2024unfolding, neyem2024exploring, 10.1145/3626252.3630960, 10893102,  10646888, ouyang2024comparing,  10420194,  chamberlain2025large, 10578869,  chen2024plagiarism, 10.1145/3708897, 10.1145/3708897, 10.1145/3649217.3653558, 10.1145/3649217.3653608, cipriano2023gpt, cipriano2024picture, crandall2023generative, crandall2024wip, grande2024student, de2024solving,  del2024automating,  dosaru2024code, doughty2024comparative, dunder2024kattis, duong2024automatic,  farinetti2024chatbot, feffer2023ai, fenu2024exploring, freire2023may, freire2023may, fwa2024experience, garcia2023exploring, gardella2024performance, 10.1145/3643795.3648375,  grevisse2024docimological, Papakostas2024rule, penney2023anticipating,  10.1145/3649165.3690111, 10.1145/3657604.3664701, 10748200,  10478015, hossain2024diffusion, 10343215,  10578742,  10892956, 10834346, sheard2024instructor,  10578820,  10834293, sheese2023patterns,  zviel2024good, zviel2024good, xiao2024preliminary,  tanay2024exploratory, styve2024developing,  singh2023exploring, singh2023exploring, scholl2024analyzing, shen2024implications, scholl2024novice, schefer2024exploring, 10.1145/3626252.3630928,   prather2024beyond, salim2024impeding, rajala2023call, oyelere2025comparative, savelka2023can, sarsa2022automatic, 10.1145/3617367, prather2023robots, prather2024widening, nizamudeen2024investigating, modran2024developing, ma2024exploring, mohamed2025hands, 10.1145/3661167.3661273, 10.1145/3716640.3716651, 10.1145/3613944.3613946,  10.1145/3501385.3543957, 10.1145/3689535.3689554, 9874678, 10568376, 10892916, kumar2025investigating, rogers2025playing, mcdanel2025designing,  margulieux2024self,  simaremare2024pair, rajala2023call,  10569684, 10893343, 10.1145/3568813.3600139, 10.1145/3649165.3690123, hou2024effects, 10.1145/3641554.3701785, 10.1145/3643795.3648379, reeves2023evaluating, rhee2024evaluation,  smith2024prompting, hou2024more, 10398297, 10633447, 10.1145/3649217.3653607, 10.1145/3636243.3636259, 10.1145/3626252.3630784, 10923780, 10578838, 10578838, 10663001,  Jamie_2025, jordan2024need, 10.1145/3626252.3630803, 10.1145/3636243.3636252, 10665172,  10398397, tang2025sphere, 10.1145/3649165.3690125,  10.1145/3626252.3630822, ma2024enhancing,  10722841,  kiesler2023exploring, manley2024examining, 10.1145/3613905.3647967, lohr2025you, 10665141,  tsai2025selfregulation, lau2023from, 10.1145/3639474.3640058, 10.1145/3545945.3569770, 10893096,  10590605,  liffiton2024codehelp, liu2024can, 10893333, 10305701, 10590238, 10590238, 10664905, 10.1145/3545945.3569830, 10500042, 10.1145/3639474.3640076,  10684637, 10.1145/3641554.3701867, 10.1145/3641554.3701867, lohr2025you, 10.1145/3641554.3701844, karnalim2024detecting, jin2024teach, 10.1145/3641554.3701864,  humble2024cheaters, 10.1145/3632620.3671103,  hoq2023detecting, 10417871, lyu2024evaluating, zastudil2023generative, zdravkova2024unveiling, 10.1145/3636243.3636245,  mailach2025ok, 10.1145/3649850, 10.5555/3715622.3715633, akcapinar2024aichatbots, ali2025automated, hasananalyzing, alkafaween2024automating,  bikangaada2024helps, doughty2024comparative, kumar2024guiding, 10.1145/3699538.3699546, Nutalapati2024coding} \\
    Applying Knowledge &      Decline in problem-solving, coding, writing skills, homogenization of code styles.
 &
      \footnotesize
      \cite{alomar2025nurturing, amoozadeh2024student, azaiz2024feedback, azoulay2025integrity, borghoff2025generative, budhiraja2024s, cai2024compat, castillo2024ethical, 10.1145/3605507.3610629, servin2024unfolding, 10.1145/3649217.3653558, cipriano2023gpt, cipriano2024picture, del2024automating, denny2024prompt, dosaru2024code, dunder2024kattis, duong2024automatic, farinetti2024chatbot, gardella2024performance, Papakostas2024rule, 10478015, hossain2024diffusion, geller2024understanding, sheese2023patterns, zviel2024good, tanay2024exploratory, styve2024developing, scholl2024novice, schefer2024exploring, salim2024impeding, oyelere2025comparative, prather2024widening, nizamudeen2024investigating, ma2024exploring, 9874678, mcdanel2025designing, simaremare2024pair, rajala2023call, 10.1007/978-3-031-35897-5_31, 10633447, 10398397, 10722841, lau2023from, 10590238, 10.1145/3639474.3640076, karnalim2024detecting, 10.1145/3641554.3701864, zastudil2023generative, hasananalyzing, doughty2024comparative, 10.1145/3699538.3699546} \\
        \rowcolor{gray!10}
 \textbf{Metacognitive}&
       &
       \\
    Disruptions to Metacognition&
      Negative impacts on students' ability to develop a plan for solving a problem, monitor the plan, and adapt the plan as necessary.  &
      \footnotesize
      \cite{ahmed2025feasibility, al2024ai, alomar2025nurturing, alves2024give, amoozadeh2024student, amoozadeh2024trust, anderson2024using, 10578934, arora2025analyzing, axelsson2024assistance, azaiz2024feedback, azoulay2025integrity, barlowe2024wip, basu2024large, bird2024faceless, borghoff2025generative, brockenbrough2024using, cai2024advancing, cai2024compat, castillo2024ethical,  10.1145/3605507.3610629, servin2024unfolding, neyem2024exploring, 10893102, 10420194, 10420194, cipriano2023gpt, cipriano2024llms, cipriano2024picture,  grande2024student, de2024solving, denny2024prompt, dosaru2024code, duong2024automatic, feffer2023ai, fenu2024exploring, garcia2023exploring, gardella2024performance,  grevisse2024docimological,  10.1145/3649165.3690111, 10.1145/3657604.3664701, 10748200,  gutierrez2024generative, 10478015, 10578742, 10892956, 10834346, sheard2024instructor,  10578820, 10834293, 10834293, sheese2023patterns, zviel2024good, xiao2024preliminary, tanay2024exploratory, styve2024developing, singh2023exploring, scholl2024analyzing,  scholl2024novice, schefer2024exploring,  prather2024beyond, rajala2023call, oyelere2025comparative, sarsa2022automatic, 10.1145/3617367,  prather2023robots, prather2024widening,  nizamudeen2024investigating, modran2024developing, ma2024exploring, 10.1145/3661167.3661273, 10.1145/3716640.3716651,  10.1145/3613944.3613946, 10.1145/3501385.3543957,  10.1145/3689535.3689554, 9874678, 10568376, kumar2025investigating, rogers2025playing, simaremare2024pair, rajala2023call, 10.1145/3568813.3600139, hou2024effects, smith2024prompting, hou2024more, soto2024enhancing, 10398297, 10633447, 10923780, 10663001,  Jamie_2025, jordan2024need, 10.1145/3626252.3630803, 10.1145/3636243.3636252,  10665172, 10398397,  10.1145/3649165.3690125, ma2024enhancing, 10722841,  kiesler2023exploring, manley2024examining, 10.1145/3613905.3647967, lohr2025you, 10852510, 10665141,  tsai2025selfregulation, lau2023from, 10.1145/3639474.3640058,  10.1145/3545945.3569770, 10893096, liffiton2024codehelp, 10893333, 10664905, 10500042, 10.1145/3639474.3640076, 10.1145/3641554.3701867, lohr2025you, karnalim2024detecting, jin2024teach, 10.1145/3641554.3701864, 10.1145/3632620.3671103, hoq2023detecting, 10417871,  lyu2024evaluating, zastudil2023generative, zdravkova2024unveiling, 10.1145/3636243.3636245, macneil2023experiences, 10.1145/3649850, akcapinar2024aichatbots, hasananalyzing,  alkafaween2024automating, bikangaada2024helps, hsin2024effect, kumar2024guiding, 10.1145/3699538.3699546} \\
    Over-reliance and trust calibration &
      Students either over-rely on GenAI or misjudge how much to trust its outputs.  &
      \footnotesize
      \cite{aerts2024feasibility, ahmed2025feasibility, alghamdi2024exploring, alomar2025nurturing, alves2024give, amoozadeh2024student,  amoozadeh2024trust,  apiola2024first, 10578934, 10578934, axelsson2024assistance, azaiz2024feedback, azoulay2025integrity,  budhiraja2024s, cai2024advancing, cai2024compat, castillo2024ethical,  servin2024unfolding, 10420194, 10578869, 10.1145/3708897, crandall2023generative, crandall2024wip, grande2024student, de2024solving, ebert2025leveraging, farinetti2024chatbot, feldman2024non, freire2023may, garcia2023exploring, 10478015, hossain2024diffusion, 10343215, 10578742, geller2024understanding, sheard2024instructor, xiao2024preliminary, styve2024developing, singh2023exploring, scholl2024novice, schefer2024exploring, prather2024beyond, rajala2023call, prather2024widening, nizamudeen2024investigating, ma2024exploring, 10.1145/3689535.3689554, 10892916, 10892916, kumar2025investigating, rogers2025playing, margulieux2024self,  rajala2023call, hou2024effects, 10.1145/3626252.3630803, ma2024enhancing, 10.1145/3613905.3647967, lohr2025you, 10665141, liu2024can, 10305701, 10.1145/3641554.3701867, lohr2025you, 10.1145/3641554.3701844, humble2024cheaters,  zastudil2023generative, 10.1145/3636243.3636245, mailach2025ok, 10.5555/3715622.3715633, ali2025automated, hasananalyzing, bikangaada2024helps, kumar2024guiding, 10.1145/3699538.3699546} \\

    \bottomrule
  \end{tabularx}
\end{table*}

\begin{table*}[t]
  \centering
  \caption{Classroom Harms identified in the SLR}
  \label{tab:classroom-harms}
  \begin{tabularx}{\textwidth}{@{}
      >{\arraybackslash}p{3cm}   
      X                                      
      >{\arraybackslash}p{9cm}  
      @{}}
    \toprule
    \textbf{Harm} & \textbf{Definition} & \textbf{Relevant Papers} \\
    \midrule
     \rowcolor{gray!10}
 \textbf{Assessment and Academic Integrity}&
       &
       \\
    Cheating \& Plagiarism &
      Includes AI-enabled cheating and academic dishonesty. &\footnotesize
      \cite{ai2024custom, 10343189, amoozadeh2024student, amoozadeh2024trust, arora2025analyzing, axelsson2024assistance, azoulay2025integrity, barlowe2024wip, basu2024large, bird2024faceless, budhiraja2024s, buffardi2024designing, castillo2024ethical, 10.1145/3605507.3610629, servin2024unfolding, 10893102, 10420194, chamberlain2025large, 10578869, chen2024plagiarism, 10.1145/3649217.3653558, cipriano2023gpt, cipriano2024llms, cipriano2024picture, crandall2023generative, grande2024student, denny2024prompt, denzler2024style, dosaru2024code, dunder2024kattis, duong2024automatic, ebert2025leveraging, 10343215, 10578742, 10892956, 10834346,  sheard2024instructor, 10578820, sheese2023patterns, zviel2024good, tanay2024exploratory, styve2024developing, singh2023exploring, shen2024implications, schefer2024exploring, prather2024beyond, salim2024impeding, sarsa2022automatic, 10.1145/3617367, prather2023robots, modran2024developing, 10892822, 10.1145/3661167.3661273, 10.1145/3613944.3613946, 10923798, 10.1145/3593342.3593360, 10.1145/3501385.3543957, 9874678, 10568376, mcdanel2025designing, Manoharan2023contract, smith2024early, 10.1145/3568813.3600139, hou2024more,  10398297, 10.1007/978-3-031-35897-5_31,  10.1145/3633287, 10923780,   Jamie_2025, 10.1145/3626252.3630803, 10.1145/3636243.3636252, 10665172, 10398397, manley2024examining, 10665141,  tsai2025selfregulation,  lau2023from, 10.1145/3545945.3569770,  10305701, 10590238, 10664905, 10500042,  10.1145/3639474.3640076, karnalim2024detecting,  humble2024cheaters, hoq2023detecting, 10417871, lyu2024evaluating, lyu2024evaluating, zastudil2023generative,  zdravkova2024unveiling, 10.1145/3649850, akcapinar2024aichatbots, hasananalyzing,  bikangaada2024helps, fan2023exploring,  hsin2024effect,  10.1145/3699538.3699546} \\
    Fairness \& Enforcement &
      Ambiguous policies, unfair accusations without proof, and inconsistent enforcement across instructors and institutions
 &\footnotesize
      \cite{alghamdi2024exploring, ali2024using, ali2024using, 10343189, azaiz2024feedback, azoulay2025integrity, buffardi2024designing, 10.1145/3605507.3610629, nguyen2024comparing, chen2024plagiarism, cipriano2023gpt, cipriano2024llms, de2024solving, denzler2024style, doughty2024comparative, dunder2024kattis, duong2024automatic, fwa2024experience, 10343215, 10343215, 10892956, 10892956, 10834346, 10578820, schefer2024exploring, prather2024beyond, salim2024impeding, salim2024impeding, oyelere2025comparative, savelka2023can, sarsa2022automatic, 10.1145/3617367, modran2024developing, 10892822, 10892822, mohamed2025hands, miranda2024enhancing, miranda2024enhancing, 10.1145/3661167.3661273, 10.1145/3593342.3593360, 10.1145/3501385.3543957, 10.1145/3501385.3543957, 9874678, 10892916, menezes2024standup, menezes2024standup, mcdanel2025designing, Manoharan2023contract, Manoharan2023contract, Manoharan2023contract, Manoharan2023contract, hou2024more, hou2024more, 10398297, 10398297, 10633447, 10.1145/3633287, 10.1145/3633287, Jamie_2025, manley2024examining, manley2024examining, manley2024examining, 10852510, lau2023from, lau2023from, 10893333, 10664905, 10.1145/3545945.3569830, 10500042, 10500042, lohr2025you, 10.1145/3641554.3701844, karnalim2024detecting, karnalim2024detecting, karnalim2024detecting, humble2024cheaters, hoq2023detecting, lyu2024evaluating, zdravkova2024unveiling, akcapinar2024aichatbots, akcapinar2024aichatbots, ali2025automated, alkafaween2024automating, fan2023exploring} \\
       \rowcolor{gray!10}
 \textbf{Equity}&
       &
       \\
    Model biases&
      Biases in GenAI responses related to language, race, and gender. &\footnotesize
      \cite{ahmed2024potentiality, amoozadeh2024trust, 10578649, barambones2024ChatGPT, bird2024faceless, castillo2024ethical, 10893102, 10893102, 10500654, 10500654, ebert2025leveraging, feldman2024non, 10578742, singh2023exploring, prather2024beyond, oyelere2025comparative, 10.1145/3716640.3716649, 10.1145/3617367, 10.1145/3617367, prather2023robots, maurat2025comparative, simaremare2024pair, 10.1145/3568813.3600139, soto2024enhancing, 10.1145/3626252.3630784, 10923780, jordan2024need, 10.1145/3626252.3630803, 10665172, 10665141, 10590605,  liffiton2024codehelp, liu2024can, humble2024cheaters, 10.1145/3632620.3671103, 10417871, 10.1145/3699538.3699546} \\
    Access \& Opportunity&
      Unequal access to AI tools or uneven distribution of benefits.&\footnotesize
      \cite{10578649, 10578649, apiola2024first, 10578934, 10578934, 10871836, naik2024Generating, castillo2024ethical, 10893102, 10578869, 10.1145/3708897,  10500654, grande2024student, deb2025enhancing, ebert2025leveraging, ebert2025leveraging, feffer2023ai, feldman2024non, fenu2024exploring,  10478015, 10578742, 10578820,  10834293, scholl2024novice, schefer2024exploring, prather2024beyond, 10.1145/3716640.3716649, 10.1145/3617367, prather2023robots, 10892822, 10568376, rogers2025playing, 10893343, 10.1145/3568813.3600139, hou2024effects, 10.1145/3643795.3648379, hou2024more, 10398297, 10.1145/3649217.3653607,  10.1145/3636243.3636259, 10.1145/3626252.3630784, 10923780, 10343052, jordan2024need, jordan2024need, 10.1145/3626252.3630803, 10.1145/3636243.3636252,
      10665172, 10665141, lau2023from, liffiton2024codehelp, liu2024can,   10684637, 10.1145/3641554.3701867, 10.1145/3632620.3671103, 10417871, lyu2024evaluating, fan2023exploring, kumar2024guiding, 10.1145/3699538.3699546} \\
        \rowcolor{gray!10}
 \textbf{Social Harms} &
       Reliance on GenAI can reduce collaboration and increase feelings of isolation.
       & \footnotesize
       ~\cite{10343189, apiola2024first, cai2024advancing, castillo2024ethical,  10.1145/3632620.3671098, 10578869, 10578869, 10500654, PETRESCU20231028, prather2024beyond, sarshartehrani2024enhancing, 9874678, simaremare2024pair, smith2024early, 10.1145/3568813.3600139, hou2024effects,  10398297, Jamie_2025,  tsai2025selfregulation, lau2023from, bikangaada2024helps}\\
     \rowcolor{gray!10}
 \textbf{Instructional and Logistical}&
    Logistical challenges, privacy concerns, and new learning objectives
      &\footnotesize
      \cite{al2024ai, alves2024give, amoozadeh2024student, 10578649, apiola2024first, arora2025analyzing, axelsson2024assistance, brockenbrough2024using, budhiraja2024s, naik2024Generating, servin2024unfolding, neyem2024exploring, 10420194, 10578869, 10.1145/3708897, cipriano2023gpt, cipriano2024llms, deb2025enhancing, denny2024prompt, denzler2024style, feldman2024non, 10478015, 10578742, tanay2024exploratory, styve2024developing, scholl2024analyzing, 10.1145/3626252.3630928,  prather2024beyond, rajala2023call, prather2024widening, modran2024developing, meza2024improving, 10.1145/3661167.3661273, kumar2025investigating, rogers2025playing, menezes2024standup, mcdanel2025designing, margulieux2024self,  rajala2023call, 10.1145/3568813.3600139, hou2024effects, hou2024more, 10398297, 10.1145/3649217.3653607, 10923780, 10343052, jordan2024need, 10.1145/3626252.3630803,  10.1145/3636243.3636252,  10665172, ma2024enhancing, lohr2025you, 10665141, tsai2025selfregulation, lau2023from, 10.1145/3639474.3640058, 10.1145/3545945.3569770, 10893096, liffiton2024codehelp, liffiton2024codehelp, liu2024can, 10893333, 10305701, 10664905, 10500042, 10684637, 10.1145/3641554.3701867, lohr2025you,  10.1145/3641554.3701844, jin2024teach, 10.1145/3641554.3701864, lyu2024evaluating, zdravkova2024unveiling, mailach2025ok, macneil2023experiences,  10.1145/3649850, hasananalyzing, doughty2024comparative, kumar2024guiding, 10.1145/3699538.3699546} \\
    \bottomrule
  \end{tabularx}
\end{table*}

\section{Positionality Statement}
\label{sec:positionality}

Members of our research team have been involved in studies that have investigated both the benefits and challenges of AI tools in educational contexts. Our prior research focused primarily on the benefits of GenAI, but also identified a few concerns. Consequently, we did not approach this review with the belief that AI tools are inherently harmful or that their use in education should be avoided. Rather, we recognize generative AI as a transformative technology that is rapidly becoming a central aspect of society, industry, and education in the future. Given this transformative impact, we also recognize the importance of understanding how this technology impacts students and how we might shape the technology and our pedagogies to better support our students. 

Our goal for conducting this review is not to discourage the use of AI tools but to identify risks and harms so they can be understood, addressed, and mitigated. We believe that a thorough assessment of these risks is essential to maximize the benefits. 

In addition to these perspectives, our team also contains a mix of junior and senior researchers, with some members who are undergraduate students. This perspective as educators, researchers, and students further informed our analysis. 


\begin{figure}
    \centering
    \includegraphics[width=0.9\linewidth]{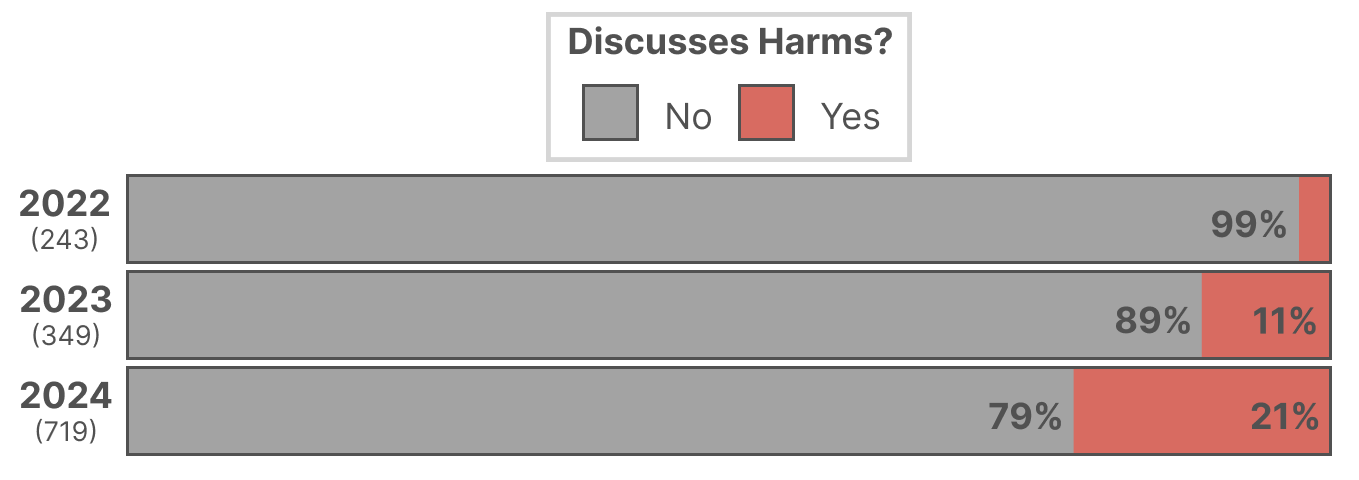}
    \caption{Stacked bar chart showing the percentage of papers that discussed harms of GenAI in computing education. Greyed bars represent papers not analyzed. \textit{Papers from 2025 and other excluded works are omitted.}}

    \label{fig:harmsbypapers}
\end{figure}

\section{Results}

As shown in Figure~\ref{fig:harmsbypapers}, there has been a consistent increase in the number of papers published about GenAI in computing education with 2024 accounting for more papers than the previous two years combined. The number of papers mentioning harms has also consistently increased from 2022, and this increase is disproportionately higher relative to the overall rise in papers discussing GenAI in computing education. This suggests that as GenAI becomes more prevalent in computing education, researchers are also becoming more aware of potential harms and pitfalls as well. 

\begin{figure}
    \centering
    \includegraphics[width=0.9\linewidth]{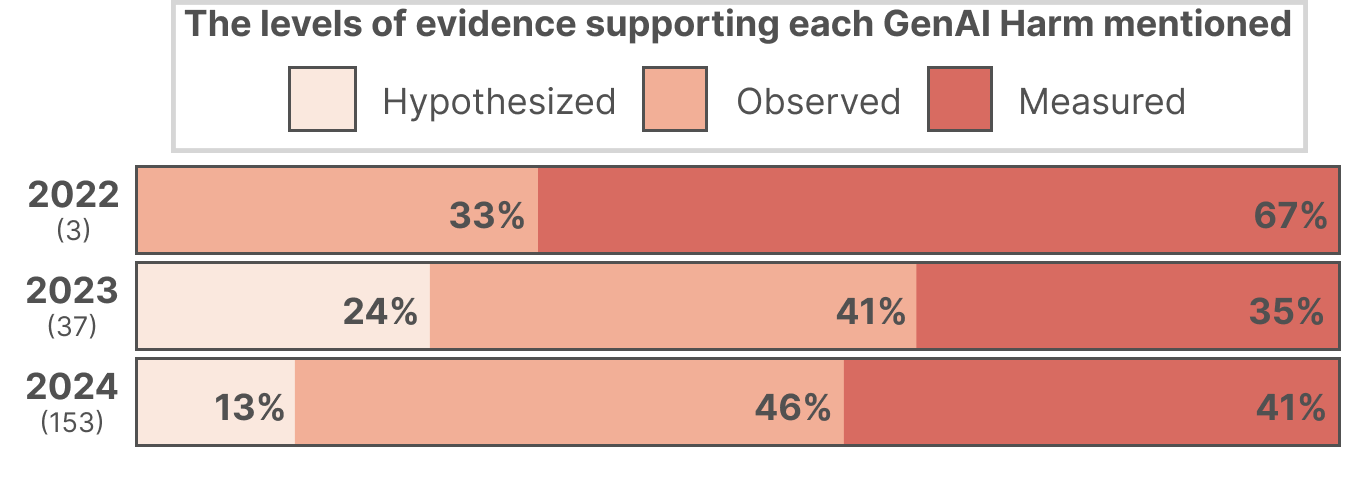}
    \caption{The stacked bar chart shows the proportion of papers based on their highest level of evidence. \textit{Papers from 2025 are omitted from this visualization}}
    \label{fig:evidenceleveloverall}
\end{figure}

However, Figure~\ref{fig:evidenceleveloverall} shows a less pronounced trend in the reporting of evidence, with only a modest increase observed between 2023 and 2024. Caution should be exercised when interpreting the 2022 data, as the sample size for that year is too small to support interpretations of trends. 



\subsection{The types and prevalence of harms}

As shown in Figure~\ref{fig:harm-table},  we identified six broad categories of harm, with multiple sub-harms. Some harms, such as those about cognition and metacognition appeared more frequently in the literature than others, such as social harms involving peer interactions and social dynamics. There were also differences in the types of evidence used to demonstrate harms. For instance, concerns about cheating and plagiarism were often supported through speculation and argumentation, rather than empirical evidence.

For less frequently mentioned harms or those with limited empirical support, this may reflect that they are indeed less prevalent or considered less important within the current research landscape. However, it may instead reflect methodological challenges for studying these harms. In the following subsections, we provide examples of the harms we identified in the literature.  

\subsubsection{Cognitive}
The most prevalent umbrella harm (32.4\%) across the papers was related to Cognition. These harms typically concerned how students acquire new knowledge or apply existing knowledge to problems. This category also had the highest proportion of evidence‑based harms (28.2\%), suggesting that concerns about cognitive impacts are not only common but also more frequently supported by empirical data.

\textbf{Acquiring Knowledge.} Of the studies that discussed Cognitive harms, most of them (86.2\%) focused on ways that GenAI harmed students' ability to acquire knowledge. Many papers reported that GenAI produces inaccurate responses, or hallucinations ~\cite{savelka2023can, hou2024more,balse2023evaluating, 10578934, bird2024faceless}, with one study noting that a third of student-AI interactions led to wrong or incomplete answers~\cite{amoozadeh2024student}. Nguyen et al. shows that as students begin to adopt GenAI as a teaching tool, its feedback can be uneven, sometimes detailed and supportive, but also occasionally inaccurate or less effective for deep learning compared with instructor feedback~\cite{nguyen2024comparing}. Students also describe the risk of shallow learning, stating that AI can make work feel `too easy', raising concerns about overdependence and reduced effort~\cite{budhiraja2024s}.

In a data structures and algorithms setting, Palacios-Alonso et al. found that when TAs guided students' prompts and verified ChatGPT's outputs, the AI-assisted group outperformed the control by 16.50 points on average~\cite{10578869}. However, for more complex algorithms (e.g, Dijkstra's, tree diagrams) outputs were sometimes inaccurate so TA checking was still required~\cite{Jamie_2025}.

Poitras et al. found that lower‑proficiency learners (reading 71.8\% vs. 78.7\%, writing 77.7\% vs. 83.9\%, both $p<.05$) relied on GenAI in 67\% of submissions primarily for generation (51\%), revision/debugging (44\%), and explanations (15\%). However, they did not outperform manual coders (no significant assignment‑score effects; $R^2=0.03$)~\cite{10.1145/3649165.3690111}. Similarly, in a controlled student experiment with Code Defenders, a gamified web platform for practicing software testing via mutants, students who relied more on the assistant produced 8.6\% fewer tests, 78.0\% of which were not useful~\cite{10.1145/3661167.3661273}.

\textbf{Applying Knowledge.} Papers that related to applying knowledge tended to focus on students getting code to run with less independent problem solving, alongside shifts in code style, reduced exploration, and weaker engagement in critical thinking. A CS2 study reported a statistically significant rise in atypical code styles after widespread adoption of ChatGPT, from 12\% of submissions in 2022 to 30\% in 2023~\cite{de2024solving}. Students also described feeling `lazy' to explore alternatives, saying that GenAI tools limited their inclination to try new ideas~\cite{simaremare2024pair}. In a semester-long field study of CodeTutor, an LLM-powered course assistant, students reported that the tool was useful for syntax help and debugging. Yet many noted it did little to foster deeper critical thinking, often encouraging reliance on quick solutions over problem analysis~\cite{lyu2024evaluating}.

A recent study also showed that in about a third of cases, students attempted to complete assignments by pasting the full task description into ChatGPT and proceeding with little or no independent work~\cite{amoozadeh2024student}. Teams sometimes adapted by decomposing complex prompts into smaller subproblems to obtain more usable outputs to address the challenges where GenAI produced unhelpful responses for harder tasks~\cite{arora2025analyzing}. 
\subsubsection{Metacognitive} The second most frequent category of harms was negative impacts on students' ability to develop a strategic plan, monitor the plan, and adapt it as needed. 

\textbf{Disruption of metacognition.} Many of the studies reviewed described the potential for disrupting students' ability to plan and to monitor their problem-solving process with quite a bit of evidence to support claims of over-reliance or over-trust in AI assistance. A paper by Scholl et al. reported that students couldn't tell if they understood responses from AI assistants or whether they were simply parroting the models~\cite{scholl2024analyzing}. In their CS1 deployment, they analyzed 213 valid student chat logs and found an average of 10.96 prompts (median 7). Shorter sessions (0–5 prompts) were dominated by direct solution requests (81\% of prompts), while longer sessions (12+ prompts) showed more mixed behaviors (41\% direct requests). This pattern reflects a shift toward “prompt-and-accept” workflows—asking for and accepting solutions—rather than engaging in planned, monitored problem solving~\cite{scholl2024analyzing}.

Related work by MacNeil et al. similarly found that students were more likely to seek direct solutions as deadlines approached~\cite{macneil2024desirable}, and Kapoor et al. reported that 50\% of students disabled guardrails to reveal full solutions, often citing time pressure or lack of self-regulation as motivations~\cite{kapoor2025exploring}.


Prather et al. also observed metacognitive challenges through an eye tracking study of how students use Github Copilot. Students with lower grades exhibited more AI-specfic difficulties and took longer to complete the task with new difficulties negatively correlated with course grade ($r=-0.503$, $p=0.020$) and positively with time to completion ($r=0.693$, $p=0.0005$). These students accepted a larger share of Copilot suggestions (34.1\% on average) than peers without metacognitive issues (24.5\%), and were more often misled or distracted by the suggestions~\cite{prather2024widening}.

Rogers et al. captured the mixed attitudes that this can create. When asked whether ``ChatGPT could hinder my ability to learn how to program for myself'', responses were bimodal, with disagreement most common (39.7\%) but agreement the second most common (25.0\%; strongly agree 8.8\%, undecided 13.2\%, strongly disagree 13.2\%)~\cite{rogers2024attitudes}.

Some papers also demonstrated that students were also concerned themselves about over-reliance, saying that: 
\begin{quote}
    \textit{``It's helping as well, but at the same time, it's reducing the human power, and human thinking ability. Like we always depend on ChatGPT for new ideas. We are not running our own minds.''}~\cite{budhiraja2024s}
\end{quote}



\noindent
\textbf{Over-reliance and trust calibration.}
Many papers describe students either over-trusting model output or leaning too heavily on it. For example, a paper by Alves and Cipriano~\cite{alves2024give} found that 47.2\% of students accepted one of the solutions provided by ChatGPT, and 72.2\% incorporated a solution fully or partially into their project code. Students also try to reduce the individual effort they put in, in a CS1 case study, about a third of attempts consisted of submitting the full task to ChatGPT with no changes~\cite{amoozadeh2024student}. Additionally, in an HCI written exam where ChatGPT use was permitted, some students reported defaulting to copying answers completely~\cite{freire2023may}.

This polarization also appears in trust calibration. One survey reports 16\% distrust, 36\% neutral, and 47\% trust (\textit{N}=253), indicating sizable groups at risk of either over‑ or under‑reliance when planning and monitoring their work~\cite{amoozadeh2024student}.

Students themselves articulate this risk. In reflections from an ethics assignment, one group warned that novices may not detect their AI mistakes and urged validating outputs against other sources ~\cite{grande2024student}. Oyelere et al. reported a positive correlation between using AI for complete solutions and the belief that knowing how to program is unnecessary ($r=0.321$, $p<0.001$) ~\cite{oyelere2025comparative}. Faculty also share these concerns, Azoulay et al. argue that the ease of AI solutions can distance independent problem solving ~\cite{azoulay2025integrity}.

In reflections around a custom tool, CodeHelp, some students admitted to going straight to the tool and feeling too dependent on it~\cite{liffiton2024codehelp}. Connections can again be drawn to Scholl et al. and their work analyzing ChatGPT logs of a CS1 course, finding students often parroted outputs from genAI rather than building upon them and understanding~\cite{scholl2024novice}.

\subsubsection{Assessment and Academic Integrity}

\begin{figure*}
    \centering
    \includegraphics[width=\linewidth]{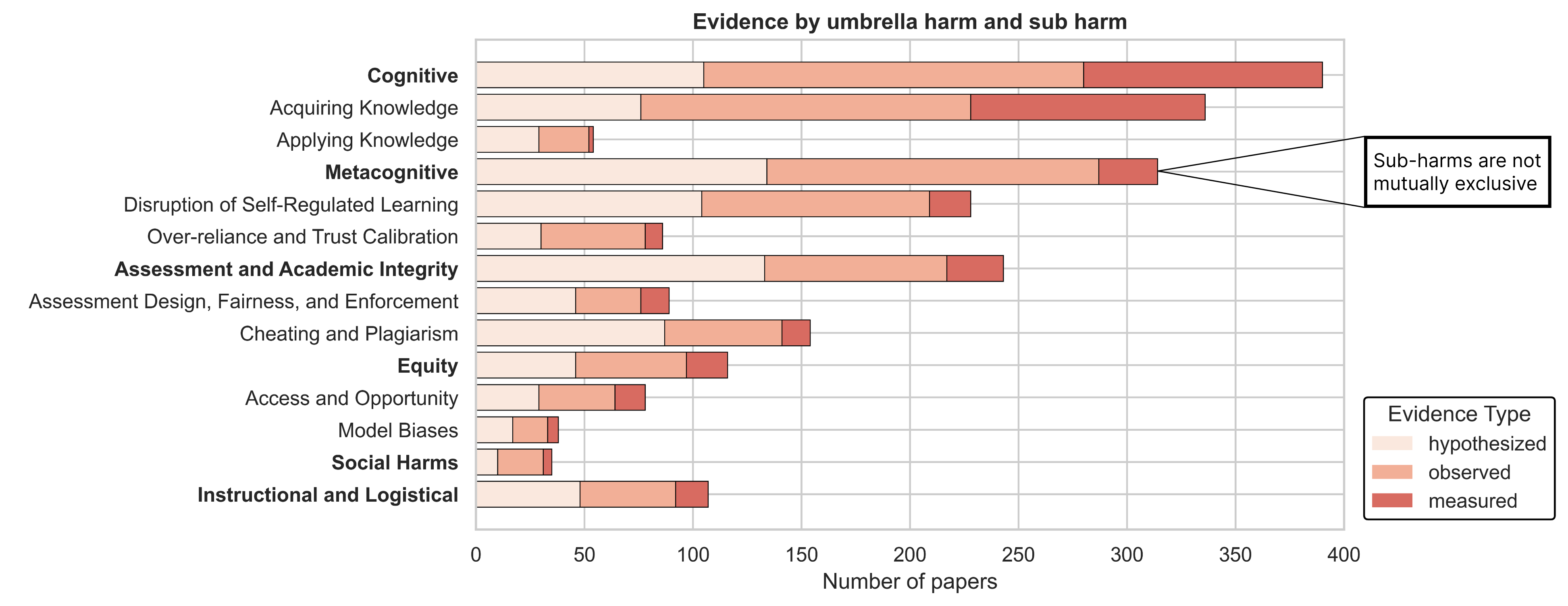}
    \caption{Umbrella harms (e.g., Metacognitive) are bolded and represent the total number of instances across all associated sub-harms. Because sub-harms are not mutually exclusive, a single paper may be counted multiple times within an umbrella category if it includes multiple sub-harms with the same type of evidence.}
    \label{fig:harm-table}
\end{figure*}

As shown in Figure~\ref{fig:harm-table}, harms related to Assessment and Academic Integrity were the third most commonly identified in our review. The most common sub-harm within this category was related to concerns about cheating and plagiarism, with a few papers also considering harms related to fairness and enforcement. 

\textbf{Cheating and plagiarism.} 
Most papers about cheating and plagiarism did not present measured evidence (8.4\%) for the harms they described. Some of the harms focused on the factors that led students to engage in academic misconduct using GenAI. For example, students tended to rely on AI when facing time pressures~\cite{axelsson2024assistance,10398397} or when they are not motivated by the work~\cite{zastudil2023generative, axelsson2024assistance, bikangaada2024helps,10.1145/3699538.3699546}. These findings largely replicate what is previously known in research on plagiarism and academic dishonesty~\cite{albluwi2019plagiarism}.  

Researchers also developed benchmarks to determine how effectively AI models could be used to solve various computer science assignments~\cite{savelka2023thrilled, hou2024more, gutierrez2024seeing, savelka2023can, reeves2023evaluating, denny2024computing, cipriano2023gpt, 10417871, cipriano2024llms, cipriano2024picture} with a framing around academic misconduct. Early work with Codex showed promising but imperfect performance: on two CS1 exams, Codex scored about 78\% on each and averaged roughly 50\% across Rainfall variants, struggling with constraints and formatting~\cite{finnie-ansley2022robots, denny2024computing}. However, subsequent papers quickly showed large gains with newer models. For example, an ITiCSE working-group replication and follow-up analysis reported GPT-4 achieving 99.5\% and 94.4\% on two CS1 exams and solving all Rainfall variants~\cite{prather2023robots, denny2024computing}. Recently, multimodal models can solve assignments, such as data structures and algorithms problems, based only on an image~\cite{hou2024more, gutierrez2024generative}.

Across these papers, it is clear that GenAI can now solve most programming problems that students encounter at various stages of their education. This quote from an interview study of faculty experiences by Sheard et al.~\cite{sheard2024instructor} summarizes it well: \textit{``I have lost my ability to confidently assess any work that students hand in.''} 

\textbf{Fairness and Enforcement.} Given these concerns about academic integrity, some of the papers discussed new considerations for assessment, including how to prevent or detect cheating, and a few discussed issues of fairness. In terms of enforcement, most papers focused on detection or prevention. The detection papers included identifying stylistic markers of AI use~\cite{denzler2024style} or the use of advanced concepts which had not been taught~\cite{neyem2024exploring}. Additional detection approaches have included code-anomaly detectors that flag unusual syntax patterns compared to student's prior work, which can help identify AI-assisted submissions. Their accuracy, however, drops in collaborative contexts~\cite{hoq2023detecting}. While promising, these methods raise fairness concerns if used as sole evidence in high-stakes enforcement.

Further studies propose preventative approaches to curb academic dishonesty, such as Prompt Problems, which present problems visually to prevent simple copy-pasting of design briefs~\cite{denny2024prompt}. Additional work evaluates  the qualities of assignments that makes them difficult to be solved directly by GenAI~\cite{mcdanel2025designing}.

Other work suggested that instead of preventing cheating, some faculty are embracing AI use~\cite{lau2023from, zastudil2023generative, 10343189, azoulay2025integrity}, acknowledging that `resistance is futile' and that students who want to cheat will continue to be able to do so~\cite{lau2023from}. 

Yet even when AI is allowed in courses, there are still unequal effects that emerge. Many students paste AI answers during assignments or exams, sometimes gaining an advantage but often being misled by incorrect responses, which raises equity concerns about who benefits and who is disadvantaged~\cite{karnalim2024detecting}.

Fairness issues also arose when AI is used on the instructional side. In blinded studies, students struggled to distinguish ChatGPT-generated feedback from that of human TAs, sometimes even accepting weaker AI feedback at face value~\cite{chen2024plagiarism}. Additional work highlights the wide spectrum of faculty enforcement stances, from outright bans to pragmatic acceptance of AI use, which raises concerns about inconsistent policies across courses and institutions~\cite{10.1145/3568813.3600138}.





\subsubsection{Equity}

Concerns about equity appeared in 29\% of papers in our review. These concerns largely focused on issues with the models or inequitable access to these models for computing students. 

\textbf{Model biases.}
Several studies described harms associated with various biases, including reports of poorer performance for low-resourced languages and the reinforcement of harmful stereotypes and prejudices.
A few papers focused on multilingual support. Prather et al. showed mixed findings with some students preferring to write code in their own native language, and others claiming that GenAI has better support for the English language~\cite{10.1145/3716640.3716649}. Jordan et al. similarly demonstrated that GenAI is more accurate in when prompted in some languages than in others~\cite{jordan2024need}.  

Biases were also present in text-to-image models. Apiola et al. taught a module to have CS students explore social justice issues in text-to-image models and found that ``the most common type of harm identified by students was the harm of application of stereotypes or prejudices''~\cite{apiola2024first}. These representational biases have the potential to reify existing biases and stereotypes that are common within computer science~\cite{sun2024smiling,cheryan2013stereotypical}.


\textbf{Access \& Opportunity.}
A few papers also identified harms related to access and opportunity. These harms included uneven usage of GenAI tools, uneven benefits, and practical barriers to participation. Early studies that investigated GenAI usage showed a bimodal distribution between students who used GenAI frequently and those who never used GenAI~\cite{hou2024effects, prather2023robots, 10.1145/3699538.3699546}. 

Beyond usage gaps, barriers include language support, subscription costs, and hardware requirements. For multilingual contexts, model accuracy and problem quality vary by language, advantaging English over some lower‑resourced languages~\cite{jordan2024need, 10.1145/3716640.3716649}. Cost and infrastructure can also exclude students or institutions~\cite{liu2024can}: for example, running OpenAI models at scale incurs nontrivial recurring costs, and local hosting requires capable GPUs. A rough estimate is \${5}–\${10} per 1,000 short conversations with GPT‑3.5 versus \${250}–\${500} with GPT‑4, and a server‑side GPU such as an RTX 3090 (24\,GB VRAM) for multi‑model deployments~\cite{liu2024can}. Beyond these institutional costs, students themselves have reported barriers such as subscription fees and device limitations that limit their ability to experiment with GenAI tools compared to peers who can afford paid access~\cite{10478015}. Faculty have also noted that licensing costs and uneven institutional resources mean some programs are unable to integrate these tools into coursework~\cite{10578869}, and others highlight that institutions without budgets for subscriptions or hardware upgrades face significant disadvantages~\cite{10500654}.

In addition to these differences in usage, Lyu et al. found that students with prior GenAI experience tended to benefit more than those without~\cite{lyu2024evaluating}. Similarly, Prather et al. demonstrated that students who were doing well in the course did even better when using Github Copilot where students who were struggling did worse~\cite{prather2024widening}. 


\subsubsection{Social Harms} Only 19 papers (8.5\%) highlighted harms of generative AI on social aspects of learning, the least frequently referenced harm in our literature review. These harms included students feeling pressured to use ChatGPT in order to keep up with their peers~\cite{10.1145/3632620.3671098}, feeling their social interactions being affected by GenAI~\cite{hou2024effects}, and feeling guilty about their AI usage~\cite{10578869}. 

Hou et al. first demonstrated the potential for social harms with examples of how students relied on ChatGPT to avoid the socio-emotional barriers of help-seeking~\cite{hou2024effects}. They included quotes from participants such as the following: 

\begin{quote}
    \textit{``It is easier for me to ask ChatGPT because I don't have to worry... I feel like it is kind of rude to continuously ask for more and more details.''}
\end{quote}

\noindent 
The following year, Arora et al. shared that three participants \textit{``noted that LLMs actually hindered collaboration due to compatibility issues between LLM-generated code from different team members''}~\cite{arora2025analyzing}, highlighting how mismatched outputs can complicate integration and slow teamwork.

   

\subsubsection{Instructional and Logistical}

Some of the studies identified ways that GenAI  created logistical challenges for instructors or required them to make substantial changes to their teaching practices. For example, in one study, students raised concerns about how to protect their data when using chatbots~\cite{rajala2023call}. Another study reported similar privacy concerns, noting that one student refused to create a ChatGPT account due to fears about data use~\cite{10500042}. In some countries, student data is protected (e.g.: FERPA in the USA) or broader protections exist for citizens more generally (e.g.: GDPR in Europe). These privacy concerns were seldom discussed.  

In some studies, concerns were also raised about the negative impacts AI has had on students' motivation. For example, students described moments where learning felt `too easy' or like `cheating,' which undermined motivation to wrestle with problems and reduced the value they placed on instructor feedback~\cite{budhiraja2024s}. At the same time, several papers pointed to an instructional gap that students can run certain tools, but often lack the skills to craft effective prompts or evaluate answers. In a semester-long study, 63\% of student prompts were rated unsatisfactory, reinforcing the need to build GenAI literacy intro courses~\cite{lyu2024evaluating}. Budhiraja et al. also noted reduced motivation alongside a need for explicit guidance on prompt engineering and effective use~\cite{budhiraja2024s}. Many expect AI to become integral to professional software practice, arguing that curricula should teach responsible and proficient use, even as some worried about job displacement~\cite{rogers2024attitudes,10.1145/3568813.3600138}. Instructors also emphasized preparing students for AI-mediated workplaces by integrating these tools into coursework~\cite{10.1145/3568813.3600138}.

Translating these needs into practice requires structure. Recent work cautions that switching to visual or proctored assessments alone does not solve the issue. AI proctoring, in particular raises accessibility, privacy, and autonomy concerns. As a result, instructors are rethinking both assessment design and how students are guided when using GenAI~\cite{hou2024more}. LLM‑based recommendations were sometimes repetitive, out of context, or futile, which students found demotivating~\cite{neyem2024exploring}. Consequently, instructors must take a more active role in scaffolding the process such as through the use of guardrails that encourage students to use tools responsibly~\cite{liffiton2024codehelp, kazemitabaar2024codeaid, macneil2025fostering, denny2024prompt}.

Finally, multiple studies described how AI is changing the relevance of various skills and competencies~\cite{10305701}. For example, instructors now may need to focus more on problem specification, code comprehension, and problem decomposition~\cite{mcdanel2025designing}. Some new skills are also required such as prompt engineering~\cite{denny2024prompt}.

\begin{figure}[htbp]
    \centering
    \includegraphics[width=\linewidth]{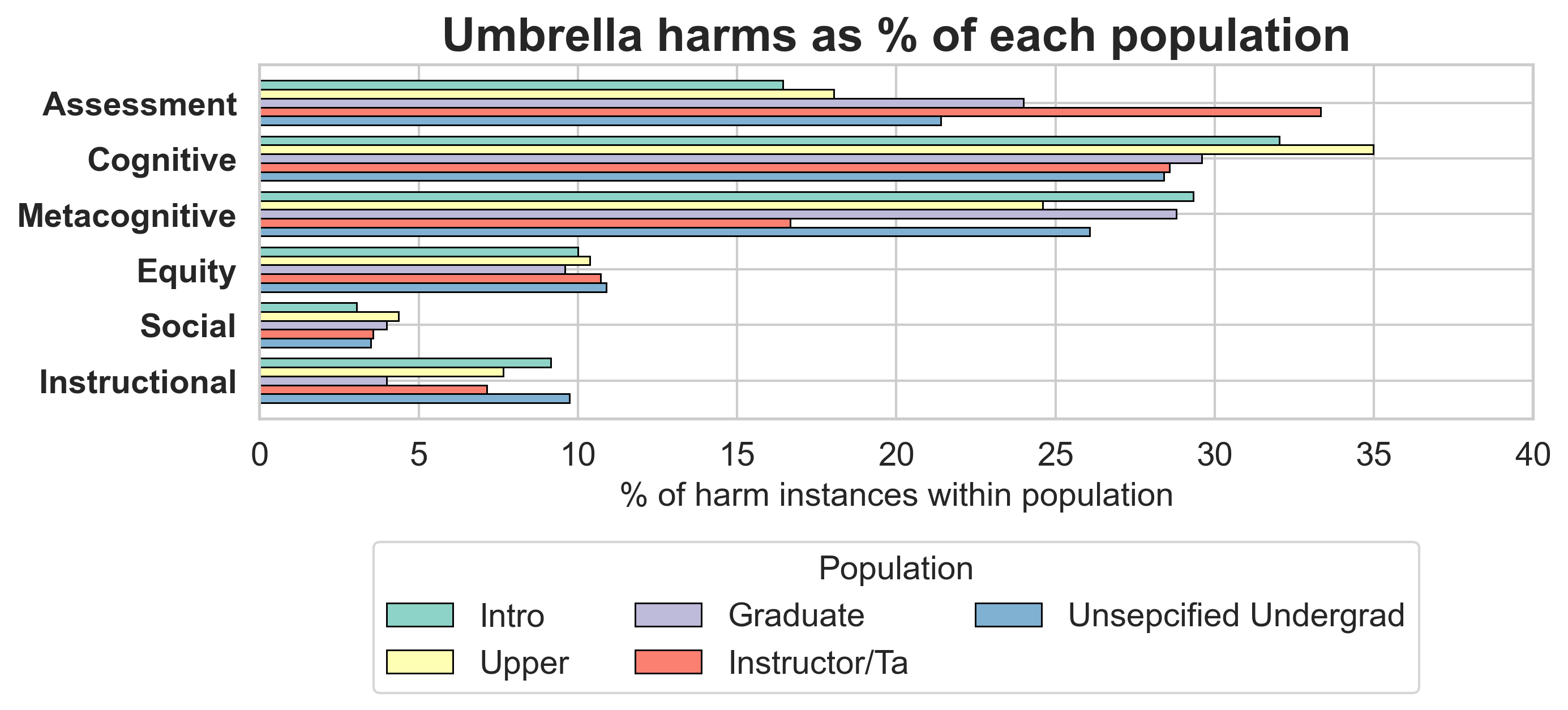}

    \caption{Harms as a percentage of each population. 69.6\% of instances involved introductory students, 12.0\% upper-level undergraduates, 9.3\% graduate students, and 9.1\% educators}  
    \label{fig:population}
\end{figure}

\subsection{The Contexts of Harms}

\subsubsection{Population}

To better understand who is most impacted by harms, we categorized harm instances by the targeted learner population. Most papers focused on introductory students, followed by unspecified undergraduate groups (where the population was only described as undergraduates), and upper-level students. Harms relating to cognitive and metacognitive challenges were most prevalent for these groups. For example, intro-level students accounted for 189 cognitive harms and 173 metacognitive harms, which was much larger than other populations.

Assessment-related harms were more evenly distributed with notable instances being reported among graduate students (n=30) and instructors/TA's (n=28). Equity harms were also cited across populations, however, they most prominently appeared among intro and unspecified undergraduate students.

\begin{figure}[H]
    \centering
    \includegraphics[width=\linewidth]{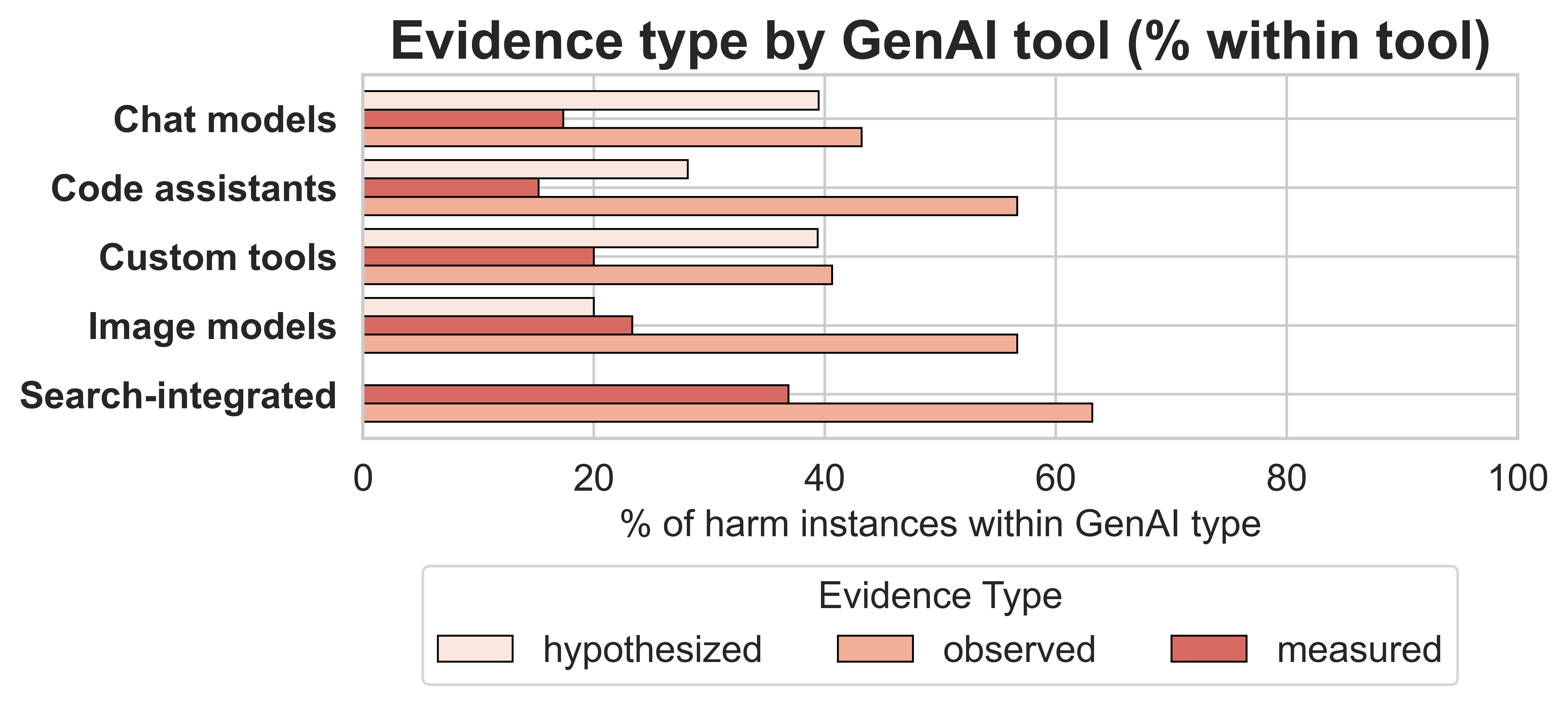}
    \caption{Percentage of harm instances by evidence type within each type of GenAI tool.}
    \label{fig:GenAItype}
\end{figure}

\subsubsection{GenAI Tool Type}

\begin{figure}[H]
    \centering
    \includegraphics[width=\linewidth]{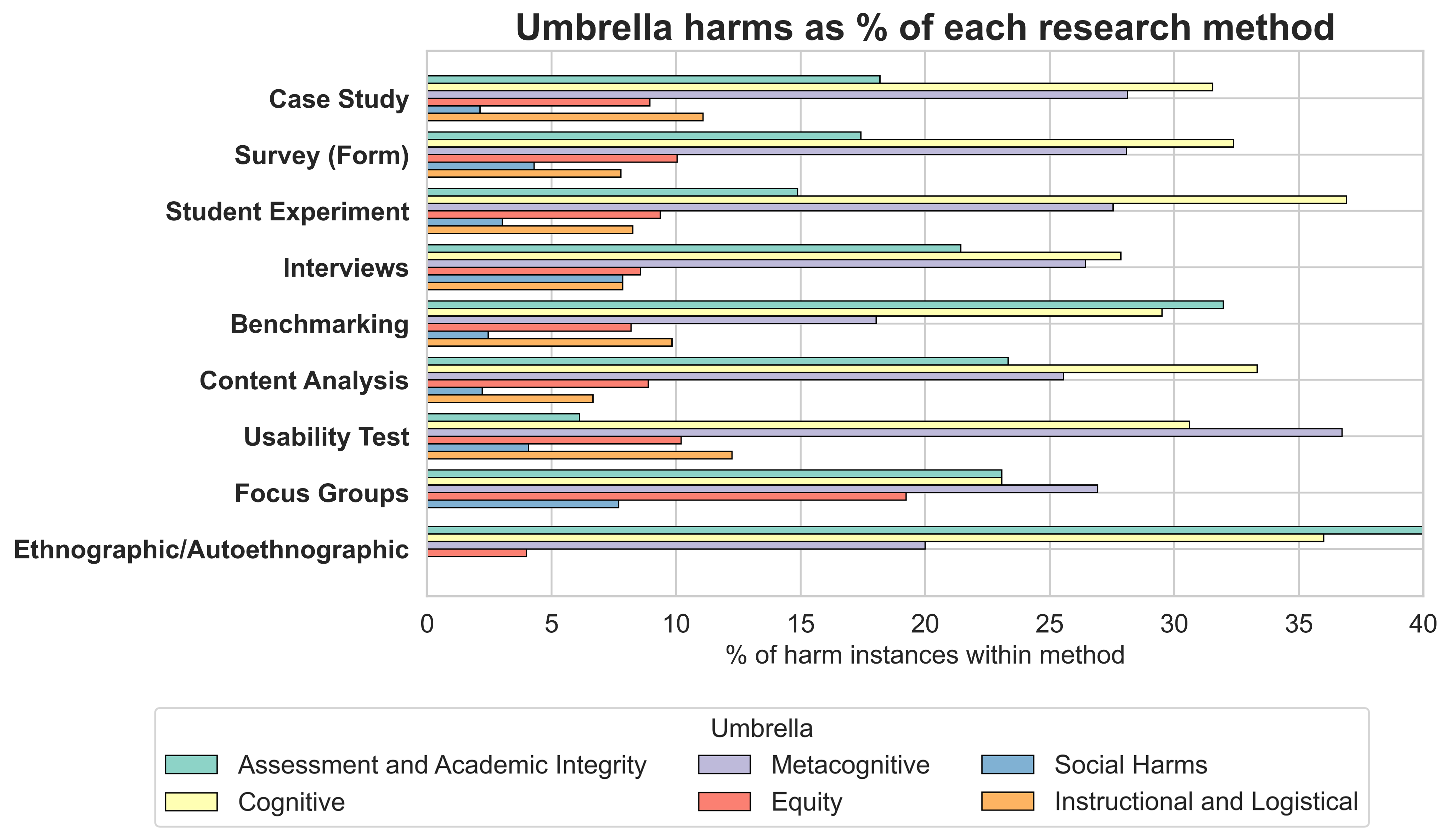}
    \caption{Umbrella harms across research methods.}
    \label{fig:method}
\end{figure}

We also examined how different GenAI tools mapped onto different harm categories Figure ~\ref{fig:GenAItype}. Chat models (eg. ChatGPT, Claude, Gemini) were the most common across all harm types, accounting for 214 instances of harms related to academic integrity, 323 cognitive, and 266 metacognitive. This most likely shows the widespread and general purpose use of chat models making them relevant for a range of student activities.

Code assistants (e.g, Github Copilot, CodeWhisperer) were the second most commonly discussed tool type. The most frequently reported harms associated with these tools were cognitive (n=79), academic integrity (n=53) and metacognitive (n=59). This aligns with their primary use case being code generation, which could impact how students think through programming tasks.

Custom or specialized tools (e.g., course-specific prototypes or research-built interfaces) were most commonly linked to cognitive (n=62) and metacognitive (n=46) harms, which could be a result of targeted experiments with these custom tools.

Image models (eg. Dall-E, Midjourney) and search-integrated tools (e.g., Perplexity, Gemini Search) appeared far less frequently. These tool types were occasionally connected to equity or social harms, but their overall harm counts were much lower, likely a result of their more limited scope in computing coursework.


\subsubsection{Research Methods}

As shown in Figure ~\ref{fig:method}, harms were not uniformly distributed across research methods. Methodological choices may have shaped the types of harms that researchers identified, as some methods may surface certain harms more readily than others. For example, cognitive harms were the most frequently reported category, appearing in 36.9\% of student experiments, 33.3\% of content analyses, and 27.9\% of interviews. Metacognitive harms were also common, most visible in surveys (28.1\%) and interviews (26.4\%). Benchmarks primarily highlighted assessment and academic-integrity harms (36.8\%).

\section{Discussion}

Our systematic literature review provides a useful and timely overview of the harms that have been hypothesized, observed, or measured by researchers and practitioners. Our results can be contrasted with a recent working group by Prather et al. where they reviewed the literature on specific GenAI interventions and systems~\cite{prather2024beyond}. In that work, they reviewed 71 papers and focused on studies that featured an explicit intervention. They found 5 studies which shared negative outcomes compared to 46 positive outcomes, 15 mixed outcomes, and 5 instances which were neutral. This review provides a much more comprehensive view, focusing on 224 papers and not just papers with instructor interventions. We found 16.3\% of the papers discussed harms compared with 7.0\% in that prior work. This may reflect a positivity bias for researchers studying their own interventions or it may be explained by a growing emphasis on negative papers. In either case, our findings show the breadth and depth of harms within computing education. 



We also observed a bias toward certain types of harms when compared with others. Cognitive and metacognitive harms were rather prevalent, but other harms such as negative social impacts of GenAI were much less prevalent. Although a few anecdotal accounts suggested that students were relying on AI instead of their peers and interfering with collaboration. A recent paper by Hou et al. provides a much more comprehensive view into the ways that AI is disrupting peer interactions and negatively impacting learning communities~\cite{hou2025all}. Another paper discusses students preferring to ask ChatGPT questions instead of asking their professors or TAs, in fear of judgment or asking a dumb question, despite acknowledging ChatGPT's inaccuracy~\cite{arora2025analyzing}. For other harms, such as equity, which also received limited attention, there was very limited discussion about the impacts on eventual careers, society, or the environment. This shows how the harms are still being understood and suggests that more research is needed to understand the harms beyond the well-studied harms related to assessment, cognition, and metacognition.






\subsection{Grand Challenges}

Our taxonomy of harms, which includes assessment, cognition, metacognition, equity, social, and logistical categories, serves as an important road map or grand challenges to guide future research. As future work, researchers should consider the following questions: 

\subsubsection{What is the future of assessment?} There appears to be growing evidence that GenAI can solve most programming problems which makes it hard to assess student outputs directly. One proposed solution has been to assess process instead of product. However, as GenAI tools increasingly support process-level tasks (e.g.: problem decomposition), this too may remain a challenge. 

\subsubsection{What cognitive harms remain as AI models improve?} While many of the cognitive harms in the papers were attributed to hallucinations or errors, these are being rapidly addressed by model improvements. However, how do we ensure students continue to build foundational problem-solving skills when AI appears to `solve' the problems for them?

\subsubsection{How can we mitigate metacognitive harms?} Perhaps one of the most concerning harms is the ``illusion of competence'' that GenAI models offer to students~\cite{prather2024widening}. Students are misled and don't realize it, but worse, they feel productive. One recent approach helps students identify misalignments between their intentions and the AI’s plan by simply offering multiple options for each code suggestion~\cite{macneil2025fostering}. How do we help students calibrate their trust and identify these misalignments?  

\subsubsection{How can we ensure everyone is learning effectively?} Multiple papers showed uneven usage and unequal benefits for different student groups. As researchers continue to understand cognitive and metacognitive harms, it becomes clearer that these harms are not experienced uniformly. Despite early promises of democratizing education, GenAI tends to benefit students who are already well-prepared, while leaving others behind.

\subsubsection{How can we preserve our learning communities?} While few papers addressed social harms, recent work has raised more serious concerns~\cite{hou2025all}. Learning is inherently social and sense of belonging has been consistently linked to retention~\cite{tinto1975dropout}. How do we preserve peer interactions, collaboration, and friendships within the class? 

\subsubsection{What new logistical challenges must we face?} GenAI has introduced a litany of new challenges for educators, including identifying new competencies, addressing data privacy concerns, scaffolding students’ use of AI tools, and adapting to changing student motivation and preparedness.

 
\vspace{8pt} 
These questions represent the grand challenges currently facing educators and are the direct result of harms students and instructors experience due to the introduction of GenAI. While there are many exciting possibilities going forward, our community should not lose sight of these challenges as well.







\section{Limitations}

Our review captures the current state of research on generative AI harms in computing education. However, this review only captures the harms that have been studied, and it is possible that students experience these harms to greater extent than the review might suggest. For example, the studies we analyzed examine harms associated with particular types of tools, but it is possible that some tools have received more attention than others. Some harms may appear more prominently in the literature not because they are more widespread or severe, but because they have been the subject of focused study. Conversely, the absence of evidence for certain harms in relation to specific tools should not be taken as evidence that such harms do not occur.

We also caution readers against interpreting our findings as suggesting that GenAI tools only harm students. While this review focuses on potential risks and negative impacts documented in the literature, GenAI also offers many opportunities to enhance student learning and engagement. Recognizing harms is a necessary step toward designing fair and effective educational experiences, but they are not a reason to reject or avoid AI altogether.

Finally, this is a scoping literature review with the goal to identify the primary types of harms and their relative prevalence. While we report on the types of evidence presented in the literature, a more rigorous investigation, such as a meta-analysis, could be a valuable future contribution to understand not just the types of harms but the magnitude of their impacts. 

\section{Conclusion}

In this paper, we present the results of a systematic literature review that scopes the harms of GenAI in computing education. By carefully analyzing 224 that were chosen based on our inclusion and exclusion criteria, we found an encouraging trend that researchers are not only focusing on the benefits but also on the harms. However, this number is still quite low with only 21\% of papers discussing harms in 2024. Our findings also show that some harms such as those related to impacts on cognition, metacognition, and assessment are more common, other harms were not as prevalent. Our work shows that despite many benefits, there is also a real potential for harm when educators do not consider the broader tool limitations or classroom contexts. 


\begin{acks}
This paper was supported by Boost Funding from the College of Science and Technology at Temple University. We thank the anonymous reviewers for their insightful and detailed feedback.
\end{acks}

\balance

\bibliographystyle{ACM-Reference-Format}
\bibliography{sample-base}

\end{document}